\documentclass[11pt]{article}

\usepackage{amsmath,amsthm, amssymb,amsfonts,subfigure}
\usepackage{epsfig,cancel}
\usepackage{dsfont}

\numberwithin{equation}{section}

\newcommand{\beq}{\begin{equation}}
\newcommand{\eeq}{\end{equation}}
\newcommand{\ba}{\begin{array}}
\newcommand{\ea}{\end{array}}
\newcommand{\bea}{\begin{eqnarray}}
\newcommand{\eea}{\end{eqnarray}}
\newcommand{\bean}{\begin{eqnarray*}}
\newcommand{\eean}{\end{eqnarray*}}
\newcommand{\eref}[1]{(\ref{#1})}

\newcommand{\comment}[1]{}

\newcommand{\cN}{{\cal N}}

\newcommand{\cA}{{\cal A}}
\newcommand{\cB}{{\cal B}}
\newcommand{\cC}{{\cal C}}

\newcommand{\cV}{{\cal V}}

\def\cjn1{{\cA, \cC^*\otimes \wedge^j \cN^*}}
\def\bjn1{{\cA, \cB^*\otimes \wedge^j \cN^*}}
\def\vjn1{{\cA, \cV^*\otimes \wedge^j \cN^*}}
\def\cjn2{{\cA, \cC\otimes \wedge^j \cN^*}}
\def\bjn2{{\cA, \cB\otimes \wedge^j \cN^*}}
\def\vjn2{{\cA, \cV\otimes \wedge^j \cN^*}}

\def\fnote#1#2{\begingroup\def\thefootnote{#1}\footnote{#2}
     \addtocounter{footnote}{-1}\endgroup}

\begin{document}

\vspace{1cm}

\title{{\huge \bf
Stabilizing All Geometric Moduli \\
in Heterotic Calabi-Yau Vacua
}}

\vspace{2cm}

\author{
Lara B. Anderson${}^{1}$,
James Gray${}^{2,3}$,
Andre Lukas${}^{4}$,
Burt Ovrut${}^{1}$
}
\date{}
\maketitle
\begin{center} {\small ${}^1${\it Department of Physics, University of
      Pennsylvania, \\ Philadelphia, PA 19104-6395, U.S.A.}\\[0.2cm]
      ${}^2${\it Arnold-Sommerfeld-Center for Theoretical Physics, \\
       Department f\"ur Physik, Ludwig-Maximilians-Universit\"at M\"unchen,\\
       Theresienstra\ss e 37, 8033 M\"unchen, Germany}\\[0.2cm]
       ${}^3${\it Max-Planck-Institut f\"ur Physik -- Theorie,\\ 
       F\"ohringer Ring 6, 80805 M\"unchen, Germany}\\[0.2cm]
       ${}^4${\it Rudolf Peierls Centre for Theoretical Physics, Oxford
       University,\\
       $~~~~~$ 1 Keble Road, Oxford, OX1 3NP, U.K.}}\\

\fnote{}{andlara@physics.upenn.edu,~James.Gray@physik.uni-muenchen.de,~lukas@physics.ox.ac.uk,~ovrut@elcapitan.hep.upenn.edu} 

\end{center}

\abstract{\noindent } We propose a scenario to stabilize all geometric
moduli -- that is, the complex structure, K\"ahler moduli and the
dilaton -- in smooth heterotic Calabi-Yau compactifications without
Neveu-Schwarz three-form flux. This is accomplished using the gauge
bundle required in any heterotic compactification, whose perturbative
effects on the moduli are combined with non-perturbative
corrections. We argue that, for appropriate gauge bundles, all complex
structure and a large number of other moduli can be perturbatively
stabilized -- in the most restrictive case, leaving only one combination
of K\"ahler moduli and the dilaton as a flat direction.  At this
stage, the remaining moduli space consists of Minkowski vacua. That
is, the perturbative superpotential vanishes in the vacuum
without the necessity to fine-tune flux. Finally, we incorporate
non-perturbative effects such as gaugino condensation and/or
instantons. These are strongly constrained by the anomalous $U(1)$
symmetries which arise from the required bundle constructions. We
present a specific example, with a consistent choice of
non-perturbative effects, where all remaining flat directions are
stabilized in an AdS vacuum.  \newpage

\tableofcontents

%
%

\section{Introduction}

In this work, we present a scenario for stabilizing the dilaton and
all geometric moduli in smooth, ${\cal{N}}=1$ supersymmetric vacua of
the heterotic string \cite{Candelas:1985en,Green:1987mn} and heterotic
M-theory \cite{Witten:1996mz,Lukas:1997fg,Lukas:1998yy,a1}.  Heterotic
compactifications to four dimensions on Calabi-Yau three-folds with
holomorphic, slope-stable vector bundles have produced
phenomenologically realistic particle physics models
\cite{burts,Bouchard:2005ag,Anderson:2009mh}, and have stimulated new
ideas in cosmology \cite{cosmo1,cosmo2,cosmo3}.  However, moduli
stabilization in this context has been more problematical\footnote{See
  \cite{Gukov:2003cy}-\cite{Brummer:2010fr} for related work,
  including stabilization mechanisms in heterotic orbifold models.}.
In type IIB string theory, moduli stabilization can be achieved with
KKLT type vacua \cite{Kachru:2003aw}. Here, one first fixes some of
the moduli, including the complex structure, using flux. The flux is
then ``tuned'' so that the perturbative superpotential in the vacuum
is very small. It follows that the fields which are not stabilized by
the flux only experience a small perturbative runaway. This can then
be balanced by non-perturbative effects to form a completely stable
vacuum. There are two problems which arise in trying to repeat this
approach in heterotic Calabi-Yau three-fold compactifications. First,
the Calabi-Yau condition appears to forbid the introduction of
topologically non-trivial Neveu-Schwarz flux to stabilize the complex
structure moduli{\footnote{However, see \cite{Cyrill} for a possible
    counterexample.}}. Second, even if one naively allows such field
strengths while retaining the Calabi-Yau geometry, the available flux
does not allow for a small vacuum value of the perturbative
superpotential -- see the Appendix for a proof of this in the large
complex structure limit. Thus, even if one can stabilize the complex
structure in this way, there is a resulting instability in the
remaining moduli which is too large to be balanced by non-perturbative
effects.

In this paper, instead of using Neveu-Schwarz flux, we will stabilize
the complex structure, as well as many of the other geometrical
moduli, using fundamental properties of the gauge field strength
present in any heterotic compactification
\cite{Sharpe:1998zu}-\cite{UsBigPaper}. These
effects are perturbative, compatible with the compactification
manifold being a Calabi-Yau three-fold, and give rise to ${\cal{N}}=1$
supersymmetric Minkowski vacua.  Because the superpotential vanishes
after perturbative stabilization, this naturally avoids a runaway
potential for the few remaining moduli. These can then be stabilized
with non-perturbative effects, {\it without the need to tune any flux
  at all}. We emphasize, however, that although the problem of tuning
flux does not arise, stabilizing moduli in our approach requires very
specific choices of vector bundles.  The relevant gauge field
strengths can be in either the hidden or visible sector, or even split
between the two. However, since it has less impact on phenomenology,
in the generic discussion in the Introduction, and when presenting an
explicit example that fixes all moduli, we locate the associated
vector bundle in the hidden sector.

\vspace{0.1cm}

Let us now discuss in more detail the perturbative moduli stabilization
mechanisms at the heart of our scenario. It is well known that there
are contributions to the four-dimensional potential of a heterotic
compactification arising from non-vanishing gauge fields in the extra
dimensions. The ten-dimensional action of heterotic theories contains
the terms
\bea \label{intro1} S = - \frac{1}{2
  \kappa_{10}^2} \frac{\alpha'}{4} \int_{{\cal M}_{10}} \sqrt{-g}
\left\{ \textnormal{tr} F^2 - \textnormal{tr} R^2 \right\}+\dots  \ . \eea 
Using an integrability condition on the Bianchi identity,
\eqref{intro1} can be rewritten, for the case of a Calabi-Yau
compactification, as  
 \bea \label{intro2}
S = -\frac{1}{2 \kappa_{10}^2} \alpha'
\int_{{\cal M}_{10}} \sqrt{-g} \left\{ -\frac{1}{2} \textnormal{tr}
  (g^{a \bar{b}} F_{a \bar{b}})^2 + \textnormal{tr} (g^{a \bar{a}}
  g^{b \bar{b}} F_{a b} F_{\bar{a} \bar{b}})\right\} +\dots \ . \eea 
The integrand in \eqref{intro2} contains no four-dimensional
indices -- $a,b$ are holomorphic and $\bar{a}, \bar{b}$ anti-holomorphic
indices with respect to a chosen complex structure on the Calabi-Yau
three-fold.  Hence, upon dimensional reduction, \eqref{intro2} gives
rise to a potential in the four-dimensional theory. For the low-energy
theory to be ${\cal N}=1$ supersymmetric it must be possible to
express the potential coming from \eqref{intro2} in terms of F- and
D-terms. Indeed, the link between supersymmetry and \eqref{intro2} is
rather direct. To preserve supersymmetry, the gauge fields in a
heterotic compactification must satisfy the Hermitian Yang-Mills
equations of zero slope; that is,
  \bea \label{hym} F_{ab}=F_{\bar{a}\bar{b}}=0
\;\;,\;\; g^{a \bar{b}} F_{a \bar{b}}=0 \ . \eea 
Clearly, if these equations are satisfied then \eqref{intro2} leads
to a vanishing potential. If, however, for some values of the moduli,
Eqs.~\eqref{hym} are not satisfied, then \eqref{intro2} gives rise to
a positive-definite potential in four dimensions. Thus, the potential
\eqref{intro2} can stabilize at least some of the moduli in a
supersymmetric, Minkowski vacuum. From the point of view of the
four-dimensional theory, the expressions $g^{a \bar{a}}g^{b \bar{b}}
F_{a b} F_{\bar{a} \bar{b}}$ and $(g^{a \bar{b}} F_{a \bar{b}})^{2}$
are associated respectively with F- and D-term contributions to the
${\cal N}=1$ potential.  In recent work
\cite{Sharpe:1998zu}-\cite{UsBigPaper},
it has been shown how to calculate these as explicit functions of the
moduli fields. This paves the way to using this potential to stabilize
moduli in heterotic models.

First, consider the requirement in \eqref{hym} that both the
holomorphic and anti-holomorphic components of the gauge field
strength must vanish to preserve supersymmetry.  This implies that the
associated vector bundle must be holomorphic with respect to a given
complex structure. It is clear, however, that this field strength need
not have zero holomorphic and anti-holomorphic components with respect
to a {\it different} complex structure. If this is the case, it
corresponds to the stabilization of some -- possibly all -- of the
complex structure moduli. Explicit examples, together with the
associated mathematical and field theoretic formalisms, were presented
in \cite{Anderson:2010mh,UsBigPaper}. It was shown that these
holomorphy ``obstructions'' are indeed related to non-vanishing
F-terms, but with an important subtlety.  There are regions of moduli
space where the scale of the potential is as large as the
compactification scale. In such regimes, the stabilized complex
structure moduli should never have been regarded as four-dimensional
fields at all -- they are fixed at a high scale. For regions of moduli
space where this scale is small, however, it was shown in
\cite{Anderson:2010mh,UsBigPaper} that these complex structure {\it
  are} fixed by F-terms.

The second condition for supersymmetry in \eqref{hym} requires the
vector bundle to have the geometrical properties of poly-stability and
vanishing slope. These properties depend on the K\"ahler moduli of the
Calabi-Yau three-fold, as can be seen from the appearance of the
metric in $g^{a \bar{b}} F_{a \bar{b}}=0$. Some bundles are only
poly-stable with slope zero for a restricted set of K\"ahler
moduli. In addition, due to the warping of the moduli across the
M-theory orbifold direction \cite{Anderson:2009nt} -- or,
equivalently, to 1-loop corrections in the weakly coupled string
\cite{Blumenhagen:2005ga} -- the last equation in \eqref{hym} also
involves the four-dimensional dilaton. In favourable cases, these
effects can stabilize combinations of the K\"ahler moduli and
dilaton. However, since neither slope nor poly-stability (nor, indeed,
holomorphy) depend on the overall size of the compactification, there
is always at least one unstabilized modulus remaining.  It was shown
in
\cite{Sharpe:1998zu}-\cite{Anderson:2009sw}
that these effects are associated with non-vanishing D-terms. As with
the F-terms, one must be careful in attributing this stabilization
mechanism to a D-term potential.  The scale of this potential is, once
again, often as large as the compactification scale. In such cases,
the stabilized dilaton and K\"ahler moduli should never have been
regarded as four-dimensional fields at all -- they are fixed at a high
scale. However, when this scale is small, it was shown in
\cite{Sharpe:1998zu}-\cite{Anderson:2009sw}
that the K\"ahler moduli and dilaton {\it are} directly fixed by
D-terms.
               
Given these mechanisms, we propose the following three stage
stabilization scenario for heterotic compactifications.

\begin{itemize}

\item {\bf Stage 1:} Choose part of the hidden sector vector bundle so that it is
  holomorphic only for an isolated locus in complex structure
  moduli space. This corresponds to F-term stabilization of the
  complex structure moduli.

\item {\bf Stage 2:} Choose the remaining part of the hidden sector
  bundle to be holomorphic for this isolated complex structure. In
  addition, construct the hidden bundle so that it is poly-stable with
  zero slope only for restricted values of the dilaton and K\"ahler
  moduli. This, we will show, is easily achieved by an appropriate
  choice of line bundles and corresponds to D-term stabilization of
  these moduli. It is possible to fix all but one of the remaining
  geometric moduli in this way. However, as we will see in stage 3,
  leaving more than one modulus unconstrained at the second stage is desirable.

\item {\bf Stage 3:} A crucial point about stages 1 and 2 is that the
  resulting moduli space of vacua is supersymmetric and
  Minkowski. That is, the unstabilized fields have no potential and
  the cosmological constant vanishes. In the final stage of our
  scenario, we fix these remaining degrees of freedom using a more
  traditional mechanism -- non-perturbative effects such as gaugino
  condensation and membrane (or string) instantons. The inclusions of
  such effects is extremely constrained. The D-terms introduced in
  stage 2 are associated with anomalous $U(1)$ symmetries under which various linear combinations of
  the axions transform. Any allowed non-perturbative superpotential must be
  consistent with these $U(1)$ symmetries.  We find this restriction sufficiently severe
  that -- if {\it only one linear combination} of the K\"ahler moduli
  and dilaton is left unstabilized in stage 2 --  it is not possible to
  fix this modulus in a controlled regime of field space. If, however,
  {\it two} moduli remain to be stabilized, then non-perturbative effects
 consistent with the $U(1)$ symmetries can fix the remaining moduli. Moreover, 
 this can be achieved in a region of moduli space where the effective field theory is valid.
 We will present an explicit example of such a vacuum.
\end{itemize}

The structure of the paper is as follows. In Section 2, we introduce
the perturbative F- and D-terms discussed above. These will be used to
carry out the first two stages of our stabilization mechanism in
Section 3. This section also includes an explicit example of stage 2
and a demonstration that the moduli can be fixed in a controlled
regime of the effective theory. In Section 4, we describe the
non-perturbative contributions to the potential. These will be used in
Section 5 to discuss the full scenario.  Finally, in Section 6, we
conclude.  In addition, a technical Appendix discussing the perturbative superpotential generated by heterotic Neveu-Schwarz flux is attached.

\section{Perturbative contributions to the potential}

In this section, we review the perturbative F- and D-term
contributions, introduced in
\cite{Sharpe:1998zu}-\cite{UsBigPaper},
to the four-dimensional potential of heterotic M-theory vacua. These
will be important in stages 1 and 2 of our moduli fixing
scenario. Specifically, the vacua we consider are smooth Calabi-Yau
compactifications of the ten-dimensional $E_8\times E_8$ heterotic
string (or its eleven-dimensional strong-coupling counterpart) with a
gauge bundle in each of the two $E_8$ sectors. These bundles are both
of the form  $V = {\cal U} \bigoplus_{I} {\cal L}_{ I}\;.$
Hence, in each sector, they consist of a non-Abelian, indecomposable
piece, ${\cal U}$, and a sum of line bundles, ${\cal L}_{
  I}$. 

\subsection{F-terms} \label{fsec}

The F-term contributions, associated with the failure of the gauge
bundles to be holomorphic, have been discussed in detail in
\cite{Anderson:2010mh,UsBigPaper}. It is sufficient for the purposes
of this paper, to illustrate our stabilization mechanism within the
context of an explicit example.

Consider the complete intersection Calabi-Yau three-fold defined by
\bea
X = \left[ \ba{c|c} \mathbb{P}^1 & 2 \\
  \mathbb{P}^1  & 2 \\
  \mathbb{P}^2 & 3 \ea \right]^{3,75}  \; . \eea 
We construct a rank 2 holomorphic bundle ${\cal U}$ on this three-fold
via the short exact ``extension'' sequence
  \bea 0 \to {\cal L} \to {\cal U} \to {\cal
  L}^* \to 0 \ , 
  \label{next1}
   \eea 
   where ${\cal L}$ is the line bundle ${\cal O}_{X}(-2,-1,2)$. At any
   point in the $75$-dimensional complex structure
   moduli space, with moduli denoted $Z^a$, the holomorphic extensions correspond to elements of
\begin{equation}
{\rm Ext}^{1}({\cal{L}}^{*},{\cal{L}})=H^1(X, {\cal L}^{2}) \ .
\label{home1}
\end{equation}
It is well-known that the dimension of a sheaf cohomology, while
possessing a generic value, can ``jump'' at special values of complex
structure. For the example discussed here, it was shown in
\cite{Anderson:2010mh,UsBigPaper} that \eqref{home1} vanishes
everywhere in complex structure moduli space {\it except} on a
specific $58$-dimensional sub-locus, where $h^1(X, {\cal
  L}^{2})=18$. The dimensions of such cohomologies are computed in
this work using techniques and code created in the development of
\cite{Anderson:2009mh,cohom}. We choose a point $Z^a_0$ on this
sub-locus and a non-vanishing extension class {\it far} from
zero. Corresponding to this choice is a holomorphic, indecomposable
$SU(2)$ bundle ${\cal U}$. Now move {\it infinitesimally} to a generic
point $Z_0^a+\delta Z^a$ {\it not} on this sub-locus. Then, $h^1(X,
{\cal L}^{2})=0$ and the only holomorphic bundle is the direct sum
${\cal{L}}\oplus {\cal{L}}^{*}$. Since an indecomposable $SU(2)$
bundle cannot split into a direct sum under an infinitesimal change in
complex structure, it is clear that ${\cal{U}}$ is not holomorphic at
a generic point in moduli space. That is, the holomorphicity of
${\cal{U}}$ is ``obstructed'' in the $75-58=17$ directions in complex
structure moduli space leading away from the special sub-locus.

As discussed in \cite{Anderson:2010mh,UsBigPaper}, these obstructions
correspond to specific non-vanishing F-terms in the effective theory
and, hence, the breakdown of supersymmetry.  It is straightforward to
determine the zero-mode spectrum of the bundle ${\cal{U}}$ defined in
\eqref{next1}. As above, consider a point $Z_0^a$ on the
sub-locus. For a non-vanishing extension class {\it far} from zero,
there are $h^{1}(X,{\cal U} \otimes {\cal{U}}^{*})=h^{1}(X, {\cal
  L}^{2})-1=17$ bundle moduli. However, to discuss the F-term
structure it is helpful to first consider bundles near $0 \in {\rm
  Ext}^{1}({\cal{L}}^{*},{\cal{L}})$.  Here, as shown in
\cite{Sharpe:1998zu,Anderson:2009nt,Anderson:2009sw}, the low-energy
gauge group is enhanced by an anomalous $U(1)$ factor and the bundle
moduli are counted by $h^1(X,{\cal L}^2)=h^1(X,{\cal L}^{*2})=18$.  We
denote these massless fields by $C_{+}^{i}$ and $C_{-}^{j}$
respectively, with the subscript $\pm$ indicating the $U(1)$ charge.
Therefore, to lowest order, the four-dimensional superpotential is
 \bea \label{mrfterm} W =
\lambda_{ij}(Z) C^i_+ C^j_- \ . 
\label{next2}
\eea 
The dimension one coefficients $\lambda_{ij}(Z)$ are
functions of the complex structure moduli $Z^a$.  The associated F-terms are
\begin{eqnarray} \label{f1}
  F_{C^{i}_{+}}=\lambda_{ij}C_{-}^{j} + K_{C^{i}_{+}}W~&,& \quad F_{C^{j}_{-}}=\lambda_{ij}C_{+}^{i} + K_{C^{j}_{-}} W, \\ \nonumber F_{Z_{\parallel}^{a}}=\frac{\partial\lambda_{ij}}{\partial{Z}_{\parallel}^a}C^i_+ C^j_- + K_{{\mathfrak z}_{\parallel}^{a}} W ~&,& \quad F_{Z_{\perp}^{a}}=\frac{\partial\lambda_{ij}}{\partial{Z}_{\perp}^a}C^i_+ C^j_- + K_{Z_{\perp}^{a}} W
\end{eqnarray}
where we have distinguished between derivatives within the
$58$-dimensional sub-locus (specified by $58$ coordinates $Z^{a}_{\parallel}$) and those leaving this sub-locus (specified by
$17$ coordinates $Z^{a}_{\perp}$). Since the fields
$C_{+}^{i}$ and $C_{-}^{j}$ are zero-modes, for $Z_{0}^{a}$ {\it on the sub-locus}, it follows that
\begin{equation}
\lambda(Z_0)_{ij} =0 \quad \Rightarrow \quad \frac{\partial
 \lambda_{ij}(Z_0)}{\partial {Z^{a}_{\parallel}}}=0 \ .
 \label{f2}
 \end{equation}

 In the next section, we show how the
 $Z_{\perp}^{a}$-dependence in the superpotential can
 stabilize the complex structure moduli to the sub-locus where
 holomorphic, {\it indecomposable} $SU(2)$ bundles exist. In
 performing this analysis we will look for supersymmetric Minkowski
 vacua for which $W$, as well as the F-terms \eqref{f1},
 vanishes. Given this we will not need to know the exact form of the
 K\"ahler potential in \eqref{f1}.

\subsection{D-terms} \label{dsec}

The low-energy gauge group arising from a bundle of the form $V =
{\cal U} \bigoplus_{I} {\cal L}_I$ necessarily includes a number of
anomalous $U(1)$ factors, one for each line bundle,
${\cal{L}}_{I}$. Associated with each anomalous $U(1)$ is a K\"ahler
moduli dependent D-term, whose form is well-known
\cite{Sharpe:1998zu}-\cite{Anderson:2009sw}. These
four-dimensional D-terms are the low energy manifestation of the
requirement that the internal bundle be poly-stable with zero
slope. Here, we simply present these D-terms, using the notation of
\cite{Anderson:2009nt,Anderson:2009sw}. Corresponding to each line
bundle, ${\cal{L}}_{I}$, they are
\bea \label{mrdterm} D_{I}^{U(1)} = f_{I} - \sum_{L \bar{M}} Q_{I}^L G_{L \bar{M}} 
C^L \bar{C}^{\bar{M}}  \ ,\eea
where $C^{L}$ are the zero-mode fields with charge $Q^{L}_{I}$
under the $I$-th $U(1)$, $G_{LM}$ is a K\"ahler metric with
positive-definite eigenvalues and
\bea f_{I} =\frac{3}{16} \frac{\epsilon_S \epsilon_R^2}{\kappa_4^2}
\frac{\mu({\cal L}_{I})}{{\cal V}} + \frac{3 \pi \epsilon_S^2
  \epsilon_R^2}{8 \kappa_4^2} \frac{\beta_i c^i_1({\cal L}_{I})}{2
  s}\label{fiterm} \eea is a dilaton and K\"ahler moduli dependent
Fayet-Iliopoulos (FI) term \cite{Anderson:2009nt,Anderson:2009sw}. The
quantities
\bea \label{defstuff} \mu ({\cal L}_{I}) = d_{ijk} c_1^i({\cal L}_{I}) t^j t^k , \qquad  {\cal V} =
\frac{1}{6} d_{ijk} t^i t^j t^k \eea
are the slope of the associated line bundle ${\cal{L}}_{I}$ and the
Calabi-Yau volume respectively. Here $t^{i}$ are the K\"ahler moduli
relative to a basis of harmonic $(1,1)$ forms $\omega_i$, with the
associated K\"ahler form given by $J=t^i\omega_i$. Furthermore, $s$ is the
real part of the dilaton. The quantities
$d_{ijk}=\int_X\omega_i\wedge\omega_j\wedge\omega_k$ are the triple
intersection numbers of the three-fold and the $\beta_i$ are the
charges on the orbifold plane where the associated line bundle is
situated. Explicitly, these charges are 
\begin{equation}
 \beta_i=\int_X({\rm ch}_2(V)-\frac{1}{2}{\rm ch}_2(TX))\wedge\omega_i\; . \label{betai}
\end{equation} 
The parameters $\epsilon_S$ and $\epsilon_R$ are given by
 \bea\label{epsilons}
\epsilon_S = \left( \frac{\kappa_{11}}{4 \pi} \right)^{2/3} \frac{2
  \pi \rho}{v^{2/3}}\;,\quad\epsilon_R = \frac{v^{1/6}}{\pi
  \rho} \;. \eea 
Here $v$ is the coordinate volume of the Calabi-Yau three-fold, $\rho$
is the coordinate length of the M-theory orbifold and $\kappa_{11}$ is
the eleven-dimensional gravitational constant. The four-dimensional
gravitational constant $\kappa_4$ can be expressed of these
11-dimensional quantities as $\kappa_4^2=\kappa_{11}^2/(2\pi\rho
v)$. In the subsequent discussion we will set $\kappa_{11}=1$ and
further, in order to simplify the FI terms~\eqref{fiterm}, choose the
coordinate parameters $\rho$ and $v$ such that
\begin{equation}
\frac{3}{16} \frac{\epsilon_S
  \epsilon_R^2}{\kappa_4^2} =\frac{3 \pi \epsilon_S^2
  \epsilon_R^2}{16 \kappa_4^2} =1  \ .
\label{hope2}
\end{equation}
Finally, for the explicit vacua discussed in this paper, we choose
each line bundle ${\cal{L}}_{I}$ such that {\it all} of the $C^L$
fields with non-vanishing charges $Q_{I}^{L}$ are absent.  Hence, the
second term in \eqref{mrdterm} will not appear.

\section{Stages 1 and 2: Minimizing the perturbative potential }

In this section, we describe the first two stages of our
scenario within the explicit context of Section 2. 
Stage 1 involves fixing the complex
structure by setting to zero the F-terms arising from superpotential \eqref{next2}. In stage 2, using the expressions given in Subsection \ref{dsec}, we
fix linear combinations of the K\"ahler moduli and the dilaton by
solving the D-flat constraints. Crucially, both steps lead to a four-dimensional supersymmetric
Minkowski vacuum. Hence, by the end of this
section, we will have achieved a perturbative stabilization of 
all but one of the geometrical moduli, with the
resulting vacuum space having a vanishing perturbative potential.

\subsection {Stage 1: Fixing the complex structure} \label{csfix}

We will demonstrate stage 1 within the context of the explicit example
presented in Subsection 2.1. First, choose the complex structure moduli
$Z^{a}_0$ to be in the $58$-dimensional sub-locus for which an indecomposable bundle ${\cal{U}}$ can be holomorphic.  Note
from \eqref{f2} that the superpotential \eqref{next2} and the first
{\it three} F-terms in \eqref{f1} always vanish.  What are the
implications of the fourth term, $F_{Z^{a}_{\perp}}$, in
\eqref{f1}? The associated potential is
\begin{equation}
V = |F_{Z^{a}_{\perp}}|^{2}= |\frac{\partial
 \lambda_{ij}(Z_0)}{\partial {Z^{a}_{\perp}}} \langle C^i_+ \rangle|^{2}| C^j_-|^{2}+\dots \ ,
\label{box1}
\end{equation}
where we suppress the multiplicative factor of $e^{K}G^{a {\bar{a}}}$
for simplicity.

Now consider a bundle ${\cal{U}}$ defined by a non-vanishing class in
${\rm Ext}^{1}({\cal{L}}^{*},{\cal{L}})$ and, hence, by $\langle C_{+}^{i}
\rangle \neq 0$. As mentioned earlier, such a bundle only has
$C_{+}^{i}$ fields as zero-modes. Hence, the $C_{-}^{j}$ fields must
have a non-vanishing mass. It then follows from \eqref{box1} that,
in contrast to Eq.~\eqref{f2},
\begin{equation}
\frac{\partial
 \lambda_{ij}(Z_0)}{\partial {Z^{a}_{\perp}}} \neq 0 \ .
 \label{box3}
 \end{equation}
 One immediate implication is
\begin{equation}
  \langle F_{Z^{a}_{\perp}} \rangle= \frac{\partial
    \lambda_{ij}(Z_0)}{\partial {Z^{a}_{\perp}}} \langle C^i_+ \rangle \langle C^j_-\rangle
  =0 \quad \Rightarrow \quad  <C^j_->=0   \ .
\label{hope3}
\end{equation}
More interestingly, now consider the potential energy obtained from
all four F-terms in \eqref{f1} evaluated at a generic point
$Z_0^{a}+\delta Z_{\perp}^{a}$ {\it not} on
the $58$-dimensional sub-locus where non-decomposable bundles ${\cal U}$ exist. Then, to quadratic order in the field
fluctuations we find, in addition to the $C_{-}^{j}$ term in
\eqref{box1}, that
\begin{equation}
  V = |\frac{\partial
    \lambda_{ij}(Z_0)}{\partial {Z^{a}_{\perp}}} \langle C^i_+ \rangle|^{2}|\delta Z_\perp^{a}|^{2}+\dots \ .
\label{box4}
\end{equation}
where a sum over index $j$ is implied. It follows from \eqref{box3} that any of the fluctuations in the complex structure away from the special sub-locus has a positive mass and, hence, 
\begin{equation}
\langle \delta Z^{a}_{\perp} \rangle = 0 \ .
\label{box5}
\end{equation}
That is, the complex structure moduli are fixed to be on the sub-locus
where an indecomposable bundle ${\cal{U}}$ can be holomorphic!

There are several things to note about the above discussion. First,
the dilaton and K\"ahler moduli have yet to appear in the
analysis. Second, the above example is somewhat special in that it is
possible to give a four-dimensional description of the stabilization
of the complex structure. In general, for the mechanism presented in
\cite{Anderson:2010mh,UsBigPaper}, this stabilization will take place
at high scale. Hence, the fixed complex structure should never have
been included as fields in the four-dimensional theory in the first
place. In such cases, one should simply write down the low-energy
${\cal N}=1$ theory without these fields present{\footnote{Indeed,
    this will even be the case in the above example if the mass term
    in equation \eqref{box4} is of the order of the compactification
    scale.}}.

Regardless, for the rest of this paper we simply assume that the
complex structure moduli have been stabilized by some appropriate
bundle in the theory. For the subsequent stages of our scenario, we
will not need to know any more information about what this bundle
actually is, other than its second Chern class and how its structure
group (times some $U(1)$ factors) is embedded in $E_{8}$. Both this
topological quantity and the group embedding are required to satisfy
certain conditions, as we will discuss below.

\subsection{Stage 2: Fixing the K\"ahler moduli and
  dilaton} \label{stage2sec} 

For simplicity, we assume in the following that there are no matter
fields $C^I$ which are charged under the anomalous $U(1)$
symmetries\footnote{The general case, including $U(1)$ charged matter
  fields, may be interesting and is compatible with our three-stage
  scenario. However, the detailed analysis is significantly more
  complicated. The D-terms~\eqref{mrdterm} now fix linear combinations
  of the T-moduli, the dilaton and the matter fields. In addition, the
  presence of matter fields typically allows for more general
  non-perturbative contributions consistent with the $U(1)$
  symmetries. This will be important for stage 3 of our scenario. We
  defer a detailed discussion of these possibilities to future
  work.}. This can be achieved by an appropriate choice of line
bundles ${\cal{L}}_{I}$ and we present an explicit example
below. Using the results in the previous section and our choice of
conventions, the $N$ D-terms are then given by
\begin{equation}
   \label{eldterm}
   D_{I}^{U(1)} = \frac{\mu ( {\cal
       L}_I)}{{\cal V}} + \frac{\beta_i c^i_1({\cal L}_I)}{s}  = c_1^i({\cal L}_I)t_i + \gamma_Is^{-1}  \ ,
 \end{equation}
 where we find it convenient to define the ``dual'' K\"ahler moduli  $t_i=\frac{1}{\cal V}d_{ijk} t^j t^k$ as well as $\gamma_I = \beta_i c^i_1({\cal L}_I)$.

 The D-term equations $D_I^{U(1)}=0$ for $I=1,\ldots , N$ form a
 linear system of equations for the $h^{1,1}(X)+1$ variables
 $(t_i,1/s)$. The system is homogeneous which means that one modulus,
 corresponding to the overall scaling of the moduli, cannot be
 fixed. Physically, this occurs because holomorphy and
 poly-stability/vanishing slope are geometrical properties which do
 not depend on the overall size of the three-fold.  Provided that all
 of the equations are linearly independent, a non-trivial solution
 requires that $N\leq h^{1,1}(X)$.

If any of the coefficients $\gamma_I$, for definiteness say $\gamma_1$, is different from zero we can proceed by solving the first equations for the dilaton $s$ in terms of the K\"ahler moduli. This leads to
\bea \label{ssol1} s= -\frac{\gamma_1}{t_i c^i_1 ( {\cal L}_1)} \ .
\eea
Substituting this into the remaining $N-1$ equations, and
    taking the Calabi-Yau volume ${\cal{V}}$ to be finite, we obtain
    the linear equations
    \bea \label{linrel}
    \left(c^i_1({\cal L}_I) - \frac{\gamma_I}{\gamma_1}
     c_1^i ( {\cal L}_1) \right)t_i =0~, \quad I=2,\dots,N \ , \eea
   which fix a number of directions in K\"ahler moduli space. In the
   most restrictive case, that is, if we start with $N=h^{1,1}(X)$
   linearly independent D-term equations, we can solve for all of the
   K\"ahler moduli in terms of the overall scaling modulus. Then, this
   scaling modulus is the only flat direction left.

   If, on the other hand, all of the coefficients $\gamma_I=0$, then the
   dilaton drops out of the D-term equations and remains a flat
   direction. In this case, the K\"ahler moduli are constrained by
\begin{equation}
 c_1^i({\cal L}_I)t_i=0\; ,
\end{equation}
and for a non-trivial solution we should have at most $N\leq
h^{1,1}(X)-1$ linearly independent such equations. In the most
restrictive case with precisely $N=h^{1,1}(X)-1$ linearly independent
equations all K\"ahler moduli can be solved for in terms of an overall
scaling modulus. Hence, we are left with two flat directions, the
scaling modulus and the dilaton.
     
As a final comment, note that the axions associated with the
stabilized combinations of s and $t^{i}$ are ``eaten'' by massive
anomalous $U(1)$ gauge bosons through the standard supersymmetric
Higgs effect \cite{Wess:1992cp}, albeit involving fields with
non-canonical kinetic terms.

\subsubsection{An example}

As we did for stage 1, we now present an explicit realization of stage
2. This example is intended as a clear example of stage 2 of our scenario and, in particular, as 
an illustration of how the dilaton can be stabilized. It should be noted that it is {\it not}
compatible with the particular example given for stage 1 of the scenario.
However, in Section 5 we will describe how to obtain a single
consistent vacuum in which stages 1 and 2 {\it can} coexist, as well
as being compatible with explicit non-perturbative contributions.

Consider the CICY three-fold \bea \label{egconf} \left(\ba{c|ccccc}
  \mathbb{P}^3 & 0&1&1&1&1 \\
  \mathbb{P}^5 & 2&1&1&1&1 \ea \right)^{2,50} \: .\eea The triple
intersection numbers are specified by $d_{111}=2$, $d_{112}=8$,
$d_{122}=12$, $d_{222}=8$. Since $h^{1,1}(X)=2$, we need to specify
two linearly independent D-terms, in the most restrictive case. We accomplish this by choosing one line bundle on each of the two orbifold fixed planes. That is, the vector bundles on the
visible and hidden planes are of the form $V_{1}={\cal{U}}_{1}\oplus
{\cal{L}}_{1}$ and $V_{2}={\cal{U}}_{2}\oplus {\cal{L}}_{2}$
respectively, where both ${\cal{U}}_{1}$ and ${\cal{U}}_{2}$ have rank
of at least two. This gives rise to two anomalous $U(1)$ factors in
the low-energy gauge group and, hence, two associated D-terms.

On the three-fold \eqref{egconf}, the line bundle ${\cal L}_1={\cal
  O}_{X}(-2,1)$ has no cohomology for a generic complex structure.
Thus it gives rise to no $C$ fields. This is also true for ${\cal
  L}_2={\cal O}_{X}(3,-2)$. In addition, any other cohomologies which
would give rise to fields charged under the two anomalous $U(1)$'s
vanish. We use these two line bundles to stabilize the dilaton and
one K\"ahler modulus in stage 2. Given these line bundles, we find
that $\gamma_1 = -2 \beta_1+\beta_2$ and $\gamma_2=-3\beta_1 +2
\beta_2$.  Now choose ${\cal U}_1$ and ${\cal{U}}_{2}$ to have second Chern
characters
\begin{equation}
\textnormal{ch}_2({\cal U}_1) = -38 \nu_1+4 \nu_2~, \quad \textnormal{ch}_2({\cal U}_2) = 15 \nu_1 -36 \nu_2 
\label{lamp3}
\end{equation}
respectively, where $\nu_i$ is a basis of harmonic four-forms dual to $\omega_i$.
It is assumed that ${\cal U}_2$ stabilizes the complex structure as in stage 1.
 In addition we find
  \begin{equation}
  \textnormal{ch}_2 ({\cal
  L}_1)= -6 \nu_1 -4 \nu_2~, \quad \textnormal{ch}_2 ({\cal L}_2)= -15 \nu_1
-20 \nu_2  
\label{lamp4}
\end{equation}
and
\begin{equation}
\textnormal{ch}_2(TX) = -c_2(TX) =
-44 \nu_1 -56 \nu_2 \ .
\label{lamp5}
\end{equation}
Combining these results gives
\begin{equation}
\beta = -22 \nu_1 + 28 \nu_{2} \ . 
\label{lamp5a}
\end{equation}
Note that the charges on the two fixed planes are equal and
opposite\footnote{Here, and in all the examples, we have, for
  simplicity, chosen vacua where no M5 branes are present.}. We define
$\beta$ to be the fixed plane charge for the locus where the line
bundle ${\cal L}_1$ is situated. This implies $\gamma_1 = 72$ and
$\gamma_2 =122$.

For this example, equations \eqref{ssol1} and \eqref{linrel} become
\bea s = - \frac{72}{3} ((t^1)^3+ 12 (t^1)^2 t^2 + 18 t^1 (t^2)^2 + 4
(t^2)^3)/(4((t^1)^2 -2 t^1 t^2 -4 (t^2)^2)) \eea and \bea -151 (t^1)^2 +122 t^1
t^2 +424 (t^2)^2 =0 \eea respectively. Note that we have expressed the dual K\"ahler moduli
$t_{i}$ in terms of $t^{i}$ using the intersection numbers presented
above. The above equations can be solved to give the relations \bea t^1 = 2.13 t^2~, \quad s = 171 t^2
\label{lamp6}
\eea between the moduli in the vacuum.  Hence, the only remaining
flat direction is the overall scaling of all three moduli.

The D-terms we have been solving are derived (in the language of the
strongly coupled theory) for small warping.  This approximation will
be valid, in our conventions, if the moduli dependent strong-coupling
parameters, given by
\begin{equation} {\hat{\epsilon}}_{S}=\frac{{\cal
      V}^{1/3}}{s}{\epsilon}_{S}~, \quad {\hat{\epsilon}}_{R}=
  \frac{s^{1/2}}{{\cal V}^{1/3}}{\epsilon}_{R}\; ,
\label{lamp1}
\end{equation}
are sufficiently small. The Calabi-Yau volume ${\cal V}$ was defined in equation
\eqref{defstuff}.
For the example in this subsection, we find 
\begin{equation}
{\hat{\epsilon}}_{S}=0.006 \ll1
\label{lamp2}
\end{equation}
and that $\epsilon_R$ may be made arbitrarily small by increasing the
size of the one remaining modulus.

A number of other consistency checks must also be satisfied. First,
the non-Abelian bundles added to each of the fixed planes must be
slope stable. A necessary condition for this is that the topological
quantities associated with those bundles satisfy the Bogomolov bound
\cite{bogomolov} for the K\"ahler moduli evaluated on each fixed
plane. We find that this is indeed the case if 1) the rank of the
non-Abelian bundle is greater than or equal to $1$ on the first fixed
plane and 2) greater than or equal to $3$ on the second plane. One
must also check that the line bundles on each fixed plane are zero
slope inside the K\"ahler cone. Working in terms of the variables
$t^i$, the two line bundles in question are zero slope on the lines of
gradient $2$ and $3/2$ respectively. The K\"ahler cone, in these
variables, is the region between the lines of slope $4$ and $2/3$, so
this test is passed as well.

Thus we have stabilized all but one linear combination of the dilaton
and K\"ahler moduli, in a supersymmetric Minkowski vacuum, in an
allowed region of field space.

\section{Non-perturbative contributions} \label{allowednp}

Non-perturbative contributions to the superpotential in our scenario
are strongly constrained by gauge invariance. To discuss this we first
introduce the complexified dilaton and K\"ahler moduli fields
$S=s+i\sigma$ and $T^i=t^i+i 2 \chi^i$, which include the axions
$\sigma$ and $\chi^i$. The D-terms in stage 2 are associated with
Green-Schwarz anomalous $U(1)$ symmetries under which these axions
transform non-trivially.  Explicitly, these transformations read
\begin{equation}
 \label{mrtrans} 
 \delta \chi^i = -\frac{3}{16} \epsilon_S
\epsilon_R^2 c_1^i ({\cal L}_I) \epsilon~, \quad  \delta \sigma
= - \frac{3}{8} \pi \epsilon_S^2 \epsilon_R^2 c_1^i ({\cal L}_I)
\beta_i \epsilon \ 
\end{equation}
for the D-terms as given in Eq.~\eqref{mrdterm}. Note that there is
one such transformation for each D-term.

To analyze non-perturbative superpotentials, we work, without loss of
generality, in the ``K\"ahler frame'' -- where the superpotential is
gauge invariant~\cite{Wess:1992cp}. Non-perturbative corrections
typically depend on linear combinations $n_iT^i+mS$ of the moduli,
where, for now, $n_i$ and $m$ are arbitrary coefficients.  A
particular non-perturbative correction which depends on such a linear
combination is allowed only if this linear combination is $U(1)$
invariant\footnote{Here we assume the absence of singlet matter
  charged under the anomalous $U(1)$ symmetries, as discussed
  earlier. If such singlet matter is present additional
  non-perturbative corrections may be allowed and the discussion
  becomes more complicated.}. From the transformations~\eqref{mrtrans}
this implies, given the conventions~\eqref{hope2}, that
\begin{equation}
 c_1^i({\cal L}_I)n_i+\gamma_Im=0\; . \label{invcond}
\end{equation}  
We note that this is precisely the same linear system of equations, in
variables $(n_i,m)$, as the D-term equations~\eqref{eldterm} which
we have used to fix linear combinations of the moduli $(t_i,s^{-1})$
in stage 2. This means that the number of linear independent
combinations $n_iT^i+mS$ on which non-perturbative effects can, in
principle, depend equals the number of flat directions left after stage
2. For this reason, there is a tension between our desire to fix as
many moduli as possible perturbatively at stage 2 and retaining enough
flexibility with non-perturbative effects.

Let us now discuss this in some more detail and ask which, if any, of
the known non-perturbative effects can co-exist with our D-terms, that
is, with our choice of gauge bundle? We begin with gaugino
condensation which is described by a non-perturbative superpotential
\bea\label{gaugeino1} W_{\textnormal{gaugino}} = A e^{-\alpha (S -
  \beta_iT^i)} \ , \eea where $A$, $\alpha$ are constants. In our
earlier language this means we have $n_i=-\beta_i$ and $m=1$. This
choice is consistent with gauge symmetry provided that all anomalous
$U(1)$ symmetries are located on the orbifold plane opposite the one
which carries the condensate. Indeed, in this case we have
$\gamma_I=c_1^i({\cal L}_I)\beta_i$ and the
conditions~\eqref{invcond} are obviously satisfied. This fact can be
easily understood from Green-Schwarz anomaly cancellation. Given that
the anomalous $U(1)$ symmetries and the condensate are on opposing
planes, no fields on the condensate plane carry $U(1)$ charge. Hence,
there is no triangle anomaly to be cancelled on this plane and,
consequently, its gauge kinetic function which appears in the exponent
of \eqref{gaugeino1} should not transform. If we have anomalous $U(1)$
symmetries on {\it both} fixed planes they will, in general, forbid
gaugino condensates from forming in any gauge group factor. However,
this can be avoided for special topological choices. For example, if
all line bundles are chosen such that $c_1^i({\cal L}_I) \beta_i=0$,
then the associated $U(1)$ symmetries do not constrain gaugino condensate
potentials at all -- on either fixed plane.

Membrane instanton superpotentials take the form  
\bea\label{membrane1}
W_{\textnormal{membrane}} = B e^{-n_i T^i} \ ,\eea 
where $B$ and $n_i$ are constants. This means we have to satisfy the conditions~\eqref{invcond} for $m=0$. If we stabilize all but one modulus at stage 2 we need at least one of the coefficients $\gamma_I$ to be non-zero. At the same time, the D-term equations~\eqref{eldterm} as well as the conditions~\eqref{invcond} have a one-dimensional common solution space which, for finite dilaton $s^{-1}\neq 0$ cannot point into the $m=0$ direction. This means that, in this case, instanton corrections are excluded. For two flat directions left at stage 2 we have two linearly independent vectors of the form $(n_i,m)$ solving the invariance conditions~\eqref{invcond}. By taking an appropriate linear combination we see that at least one type of instanton correction is allowed in this case. 

Given these facts we should first think about the ``maximal''
stabilization scenario where we only leave one flat direction at stage
2. As argued above, there is no instanton superpotential in this
case. However, if we locate all anomalous $U(1)$ symmetries on one
orbifold plane, then gaugino condensates can form on the opposite
plane so that we can attempt to stabilize the one remaining modulus by
a race-track potential. Unfortunately, this obvious course of action
runs into a serious problem. In this case, the solution to the
invariance conditions~\eqref{invcond} is $n_i=-\beta_i$ and $m=1$ and,
hence, the D-term equations~\eqref{eldterm} are solved by
$t_i=k\beta_i$ and $s^{-1}=k$ with an arbitrary constant $k$. Hence,
the ratio of the one-loop term $\beta_it^i$ in the gauge kinetic
function relative to the tree-level part $s$ is given by
\begin{equation}
 \frac{\beta_it^i}{s}=6\; .
\end{equation} 
This means that the expansions defining our four-dimensional theory
have broken down and we can not trust any resulting vacuum. For this
reason, we will consider models with {\it two} flat directions left at stage
2 in the subsequent discussion.

\section{Stages 1, 2 and 3: Minimizing the full potential}

In this section, we combine stages 1 and 2, outlined in Subsections
\ref{csfix} and \ref{stage2sec} above, with a {\it third stage}, involving the
non-perturbative effects discussed in Section \ref{allowednp}, to give a
complete description of our moduli stabilization scenario. Making the
various stages of stabilization compatible is non-trivial. We begin by separating off stage 1.
That is, we show that it is possible to stabilize the complex
structure using only the perturbative potential described in Subsection
\ref{csfix} and, having done so, that we can simply ignore these moduli in the remaining
discussion. That this can be done is non-trivial, since there is no separation
in scale between the perturbative F-terms of stage 1 and the D-terms used in stage 2.

Once the complex structure has been fixed, we move on to stages 2 and
3 and stabilize the remaining moduli.  As we have seen, the allowed
non-perturbative effects are restricted by the presence of the
D-terms. Conversely, in order to have a stable minimum of the
potential, one can view the D-terms one can include as being restricted by the
non-perturbative effects. In Subsections \ref{nogo} and
\ref{avoidnogo}, we will describe how to fit these competing effects
together. We then finish this section by providing an explicit example
of our stabilization scenario.

\subsection{Separating off Stage 1}

We want to extremize the potential of the theory, including all
perturbative and non-perturbative effects, with respect to all fields
in the problem.  Furthermore, to preserve supersymmetry in the vacuum,
we set all F-terms and D-terms to zero. In general this means that, in
considering the stabilization of the complex structure in stage 1, one
should include contributions to the F-terms coming from the
non-perturbative effects introduced in stage 3. Since fixing these
moduli involves solving $F_{Z^a} =0$, this would modify the simple
perturbative analysis performed in Subsection
\ref{csfix}. Furthermore, the expectation values for the complex
structure moduli must be substituted into the remaining F-terms
equations which are solved in stages 2 and 3 to fix some of the
remaining fields.  Since the $F_{Z^a}$ depend on $S$ and $T^i$, so
will the solutions for $Z^a$. Thus, substituting these expectation
values back into the other F-terms introduces additional $S$ and $T^i$
dependence, which must be taken into account in the remaining
analysis.

This effect could, in principle, link perturbative and non-perturbative contributions to the potential in a complicated way. Happily, however, this is not the case for the smooth heterotic vacua discussed in this paper, as we now explain. First, a few facts. 
\begin{itemize}
\item The superpotential contains two types of
  contributions -- perturbative and non-perturbative. In our theory, these are given by
  \bea W =
  W^{({\rm P})}(Z) + W^{({\rm NP})}(Z,S, T^i) \ .
  \label{bird1} 
  \eea 
  The perturbative term, as was described in Section \ref{fsec}, does
  not depend on $S$ or $T^i$. We emphasize that this is {\it not}
  generically the case in string vacua. It arises in our theory
  precisely because our complex structure is fixed to lie in the
  image of the Atiyah map discussed in \cite{Anderson:2010mh,UsBigPaper}.  The
  non-perturbative term, which contains all fields, is much smaller
  than the perturbative contribution in any controlled regime of field
  space.
\item The K\"ahler potential takes the form
 \bea K = K_{\rm CS}(Z) + K_{\rm ST}(S, T^i) \ .
 \label{bird2}
 \eea 
 As with the superpotential, there are both perturbative and non-perturbative contributions to $K$. However, the non-perturbative contributions to the K\"ahler potential are always of higher order in our analysis and, hence, we ignore them in \eqref{bird2}. 
\item Using \eqref{bird1} and \eqref{bird2}, it follows that $F_{Z^a}$ is of the form
\bea F_{Z^a} =
F_{Z^a}^{({\rm P})}(Z) + F_{Z^a}^{({\rm NP})}(Z,S, T^i)\;. 
\label{bird3}
 \eea 
\end{itemize}

The discussion of Section \ref{csfix} was concerned with finding a
solution to $F_{Z^a}^{({\rm P})}=0$, that is, the vanishing of the
{\it perturbative} F-term. This resulted in a solution $Z^a=Z^a_0$, which is
independent of the $S$ and $T^i$ moduli. The addition of a small correction $F_{Z^a}^{({\rm NP})}$
to this F-term changes this analysis by inducing a
similarly small correction $Z^a = Z^a_0 + \delta Z^a$.
The crucial point is that, in our theory, if we substitute this perturbed solution for $Z^a$ into the other F-terms and solve for the remaining fields, then it is easy to show that the correction $\delta Z^a$ only enters into terms which are
{\it second} order in the small non-perturbative quantities.
This is due to two important features of our theory; 1) the property that $W^{({\rm P})}$ in \eqref{bird1} depends on the complex structure only and 2) the fact the analysis of Section
\ref{csfix} resulted in a supersymmetric {\it Minkowski} vacuum with
\bea
W^{({\rm P})}(Z_0)= \partial W^{({\rm P})} (Z_0) =0 \ .
\label{bird4}
\eea Hence, to achieve a result accurate to {\it first} order in small
quantities, one need only set $Z^a=Z_0^a$. One can then also forget
about the perturbative superpotential in the remaining analysis, as
this vanishes for this value of the moduli. This is what we will do in
the remainder of the paper.

This establishes a separation between stage 1 and the remaining two
stages. In the following, we will assume that the vector bundles are
chosen so that stage 1 is accomplished.  Recall that -- in each $E_8$
sector -- the vector bundle is of the form $V = {\cal U} \bigoplus_{I}
{\cal L}_{I}$. The relevant quantity in stage 1 is the subbundle
${\cal{U}}$ which, via the {\it perturbative} superpotential $W^{({\rm
    P})}(Z)$, stabilizes the complex structure moduli which can be
integrated out and, henceforth, ignored.  That is, for stages $2$ and
$3$ only the Abelian subbundles $\bigoplus_{I} {\cal L}_{I}$ with
$I=1,\dots,N$ are relevant. However, certain topological data
associated with the full bundles $V$ still appears in stages $2$ and
$3$. Before continuing, we list this data. The bundles and their
constituents must be consistent with
\begin{itemize}
\item Anomaly cancellation: ${\rm ch}_2(TX)={\rm ch}_2(V_1)+{\rm ch}_2(V_2)$
\item Bogomolov bound: \newline
 $\int_X\left(2\,\rm{rk}({\cal{U}})c_2({\cal{U}})-(\rm{rk}({\cal{U}})-1)c_1^{2}({\cal{U}})\right)\wedge
  J \geq 0$
\end{itemize}
Furthermore, the charges $\beta_i$ given by Eq.~\eqref{betai} depend on
the choice of bundle ${\cal U}$ at stage 1 and should be consistent
with the values used at later stages. Lastly the rank and
embedding of the hidden sector bundle within $E_8$ must be compatible
with the existence of the gaugino condensates which will be employed in
stage 3.  With this in hand, we continue to the full stabilization
scenario.

\subsection{Stabilizing the remaining moduli: Stages 2 and 3}

In the rest of this section we carry out stages 2 and 3 of our
scenario simultaneously, thus stabilizing the remaining geometrical
moduli in a supersymmetric vacuum. We will see that, by
allowing the two effects -- D-terms and non-perturbative F-terms -- to
coexist, one places considerable constraint on which theories can be
considered. Not only does the presence of D-terms restrict the
non-perturbative effects one can use, but the
non-perturbative potential, together with the requirement that there exist
a stable supersymmetric vacuum, restricts the form of the D-terms
in stage 2. In particular, we begin by showing that no
supersymmetric vacua exist unless the gauge bundle, and thus the D-terms,
satisfy specific constraints. When these constraints are satisfied,
however, we will find explicit supersymmetric AdS vacua with all of
the geometric moduli stabilized at a minimum in a controlled regime of
field space.

\subsubsection{A no-go result} \label{nogo}

Previously, we have seen that leaving only one flat direction after
stage 2 leads to a break-down of the expansions defining the four-dimensional 
heterotic theory. Here, we present an independent reason for why
leaving only one modulus un-stabilized after perturbative effects is
problematic. Recall that in this case, at least one of the
coefficients $\gamma_I$, say $\gamma_1$, is different from zero so
that the associated D-term equation can be solved for the
dilaton. This results in \bea \label{ssol} s = - \frac{\gamma_{1}}{ t
  _i c_1^i({\cal L}_{1})} \ .  \eea Following Section \ref{allowednp},
one can write the most general non-perturbative superpotential as
\bea \label{mrw} W = \sum_a A_a e^{-\alpha_a (S- \beta_i T^i)} +
\sum_x B_x e^{-n^x_i T^i} \ ,\eea where $n^x_i, A_a, B_x$, $\alpha_a$
are constants. To ensure gauge invariance of the instanton terms under
the first $U(1)$ symmetry, we require that
\begin{equation}
 n^x_i c_1^i({\cal L}_1)=0 \label{cond1}
\end{equation} 
for all $x$. Some of the constants $A_a$, $B_x$ may be set to zero if required for invariance under all $U(1)$ symmetries. The corresponding F-terms are
\bea \label{dilf} F_S &=& - \sum_a A_a
\alpha_a e^{-\alpha_a (S - \beta_i T^i)} -
\frac{1}{\kappa_4^2} \frac{1}{2 s} W  \\ \label{kilf}
F_{T^j} &=& \sum_a A_a \alpha_a \beta_j e^{-\alpha_a (S-  \beta_i T^i)} - \sum_x B_x
n^x_j  e^{-n^x_i T^i} + K_{T^j} W 
\eea
Multiplying Eq.~\eqref{kilf} by $c_{1}^{j}({\cal{L}}_{1})$ and using $\gamma_{1}=c_1^j({\cal L}_1) \beta_j $, Eq.~\eqref{cond1} and $K_{T^j} = - \frac{t_j}{4
  \kappa_4^2}$, we find
  \begin{equation}
c_1^j({\cal L}_1)
F_{T^j} = \sum_a A_a \alpha_a \gamma_{1} e^{-\alpha_a (S-  \beta_i T^i)} -\frac{1}{4\kappa_{4}^{2}} 
t_jc_1^j({\cal L}_1)W \ .
  \label{snow1}
  \end{equation}
  Substituting Eq.~\eqref{dilf} into
  \eqref{snow1} and setting $c_1^j({\cal L}_1) F_{T^j} =0$, we obtain
  \bea 
  W\left(\gamma_{1}+\frac{s}{2}t_j c_1^j({\cal L}_1)\right) =0 \ . 
   \label{nogo2}
  \eea 
There are now two possibilities. If $W=0$ then we are considering Minkowski vacua. Such vacua, while desirable, require a careful tuning of the constants $A_a$, $B_x$. At present we cannot justify this from string theory so we will focus on the case where $W\neq 0$ which leads to AdS vacua. Then, Eq.~\eqref{nogo2} implies that
\bea \label{snow2} 
 s = -2\frac{\gamma_{1}}{t_i  c_1^i({\cal L}_{1})} \ ,
\eea
which is clearly inconsistent with the D-flat condition
\eqref{ssol}.  We conclude that if any of the anomalous $U(1)$
factors have $c_1^i ({\cal L}) \beta_i \neq 0$, it is not possible
to simultaneously solve the D- and F-flat conditions and, hence, no
supersymmetric AdS vacua exist.

\subsubsection{Avoiding the no-go result} \label{avoidnogo}

The no-go result of the previous subsection tells us that, if we are
to successfully combine the stabilization mechanisms in stages 2 and
3, we must constrain the gauge bundle such that, for each anomalous $U(1)$,
 \bea \label{constr} c_1^i({\cal L}_{I}) \beta_i=0 \ .
\eea 
It follows from \eqref{eldterm} that the dilaton no longer appears in
any D-term. Hence, when combining the various effects in our scenario,
one can not use the full power of stage 2 to stabilize the dilaton in
linear combination with the K\"ahler moduli. It follows that one need
only include $N=h^{1,1}-1$ D-terms in the four-dimensional theory
which will stabilize an equivalent number of K\"ahler moduli. The
overall K\"ahler modulus as well as the dilaton will remain as flat
directions. Non-perturbative effects prevent us from making
``optimal'' use of the D-term stabilization at stage 2 which would
only leave one flat direction.

From Eqs.~\eqref{eldterm} and \eqref{constr}, the D-term equations $D_{I}^{U(1)}=0$ now take the form
\begin{equation}
 c_1^i({\cal L}_I) t_i=0 \, .
\label{snow3}
\end{equation}
These equations are obviously solved by choosing $t_i \propto {\beta_i}$. We take the superpotential to be of the general form~\eqref{mrw}. Recall that the gaugino condensation part is automatically gauge-invariant thanks to the condition~\eqref{snow3} while for the instanton corrections we have to impose Eq.~\eqref{invcond}. For the present case, this along with \eqref{constr} implies that $n^{x}_{i}=b^{x} \beta_{i}$ for each $x$. Then, the associated
F-terms are
\bea \label{sfts} F_S &=& -\sum_a \alpha_a A_a e^{-\alpha_a (S- \beta_i T^i)} - \frac{1}{2 \kappa_4^2 s} W
\\ \label{kfts} F_{T^j} &=& \sum_a A_a \alpha_a \beta_j
e^{- \alpha_a (S- \beta_i T^i)} - \sum_x B_x b^x
\beta_j e^{-b^x \beta_i T^i} -\frac{3}{2} \frac{1}{\kappa_4^2}
\frac{\beta_j}{\beta_i t^i} W \eea 
for $j=1,\ldots, h^{1,1}$. In Eq.~\eqref{kfts} we have used the
relation $K_{T^j} = -\frac{3}{2}
\frac{1}{\kappa_4^2}\frac{\beta_j}{\beta_i t^i}$ which follows from
$t_i \propto {\beta_i}$.  Note that every term in $F_{T^{j}}$ is
proportional to $\beta_j$. Therefore, setting all of the K\"ahler
moduli F-terms to zero leads to just one equation. We will look for
solutions to our theory where the axion expectation values appearing
in the F-terms vanish. For such a choice, we see that this equation
and $F_{S}=0$ only depend on two variables, $s$ and $\beta_i
t^i$. Note that the latter is proportional to the volume of the
Calabi-Yau three-fold, that is, $\beta_i t^i \propto {\cal V}$ since
$t_i \propto {\beta_i}$. Thus, we end up with two
constraints on two real variables from the F-terms. Recalling that the
$h^{1,1}-1$ D-terms constrain the remaining variables, one expects to
find isolated solutions to this system. This is indeed the case, as we
now demonstrate with an explicit example.

\subsection{An example}

Let us consider an example where $h^{1,1}=2$ and, hence, we need only
one line bundle ${\cal{L}}$. Furthermore, take the {\it moduli fixing}
bundle $V={\cal{U}} \bigoplus {\cal L}$ to be located in the {\it
  hidden sector}. As discussed above, the subbundle ${\cal{U}}$ is
assumed to fix the complex structure moduli and does not enter the
rest of the calculation. Now demand that there be {\it two} gaugino
condensates and a {\it single} membrane instanton present.  Note that,
although the higher rank subbundle does not enter the remaining
calculation, the condition that there be two gaugino condensates
requires that the structure group of ${\cal{U}} \bigoplus {\cal L}$ be
embedded in $E_{8}$ in such a way that the commutant has {\it two}
non-Abelian gauge factors. This is easily accomplished.  We will
specify a Calabi-Yau three-fold and the line bundle
${\cal{L}}$ shortly. However, one can get a surprisingly long way in
the analysis without giving this data, as we now show.

Although physically the parameters in the superpotential would be
determined by fundamental theory, and one would then solve for the
field values at the minimum, it is simpler in practice to proceed in
the inverse fashion. That is, we can ask what parameter values are
required in the superpotential to give a minimum with specified vacuum
expectation values for the fields. Setting the F-terms \eqref{sfts}
and \eqref{kfts} to zero for the case at hand gives us the following
result.
 \bea \label{paramsol1}
A_1 &=& B e^{\alpha_1 (s-\beta_i t^i) - b \beta_i t^i} \left(\frac{b \beta_i t^i + \alpha_2 (\beta_i t^i + s (3 +2b \beta_i t^i))}{(\alpha_1 - \alpha_2)(3 s+\beta_i t^i)}\right) \\ \label{paramsol2}
A_2 &=& - B e^{\alpha_2 (s-\beta_i t^i) -b\beta_i t^i}\left( \frac{b \beta_i
  t^i + \alpha_1(\beta_i t^i + s(3 + 2b \beta_i t^i))}{(\alpha_1
  -\alpha_2)(3 s+\beta_i t^i)}\right) \eea 
Note that the fields that appear in the analysis of the F-terms are
exactly those not constrained by the D-term. More precisely, the
dilaton, $s$, does not appear in the D-term since $\beta_i c_1^i ({\cal
  F})=0$. In addition, the D-term constrains a different combination
of K\"ahler moduli than $\beta_i t^i$.
If, for example, we ask that the dilaton be stabilized at $s=1000$ and the
overall volume be fixed at $\beta_i t^i =100$, we find the following
values solve equations \eqref{paramsol1} and \eqref{paramsol2},
\begin{equation}
A_1=-299, ~A_2 =734,~ \alpha_1 = 1/10,~ \alpha_2=10/99,~ b=1,~B=1000 \ . 
\label{desk1}
\end{equation}
Note that these are reasonable parameter choices and that the moduli
are stabilized in controlled regions of field space. Also note that
the two exponents associated with the gaugino condensates are quite
close in value. This is as expected since the dilaton here is being
stabilized essentially by the racetrack mechanism
\cite{Dixon,Krasnikov:1987jj,Casas:1990qi,deCarlos:1992da}.

\vspace{0.1cm}

\begin{figure}[!h]
\centering
\mbox{\subfigure{\includegraphics[width=3in,height=2in]{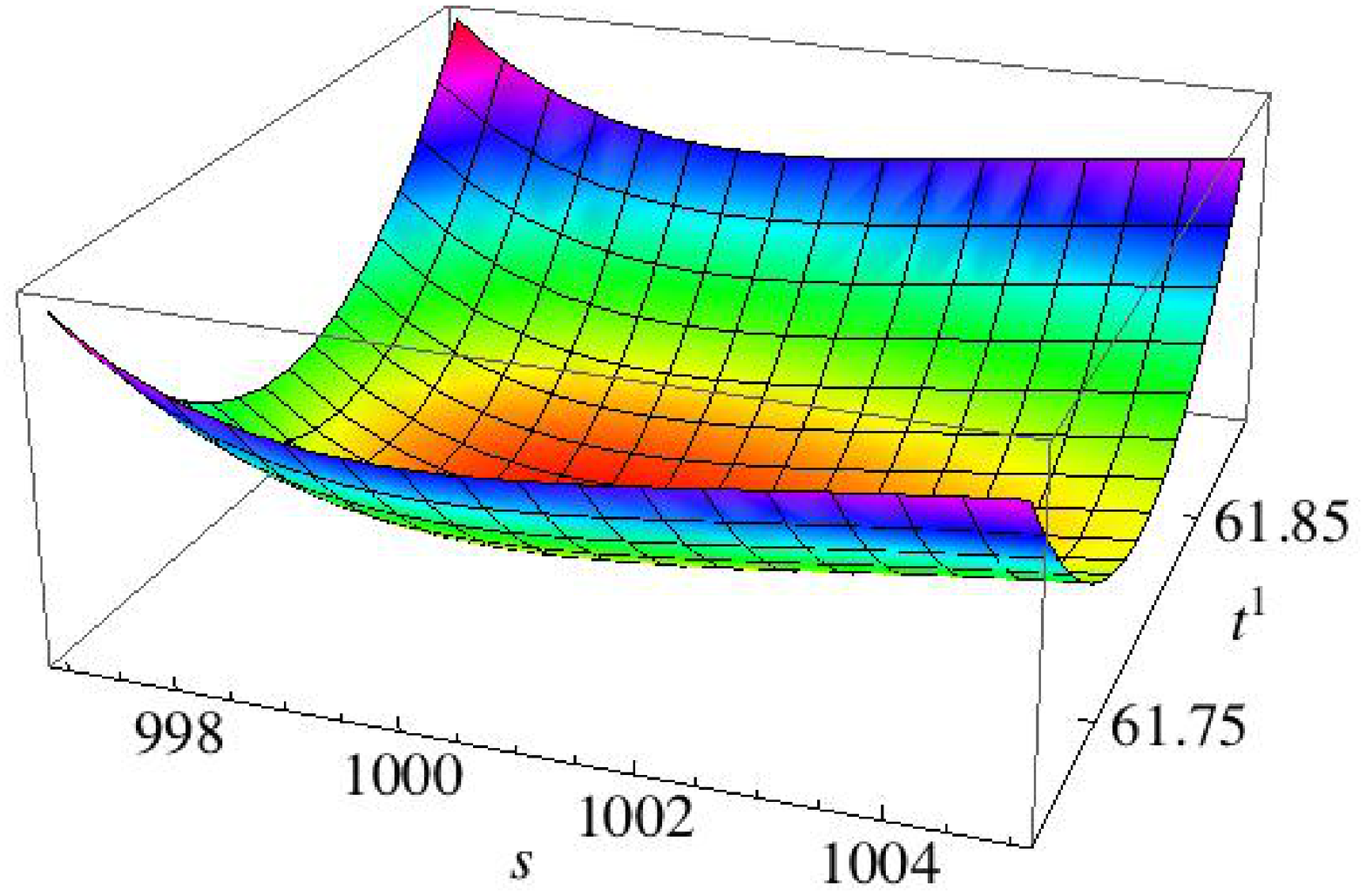}}\quad
\subfigure{\includegraphics[width=3in,height=1.9in]{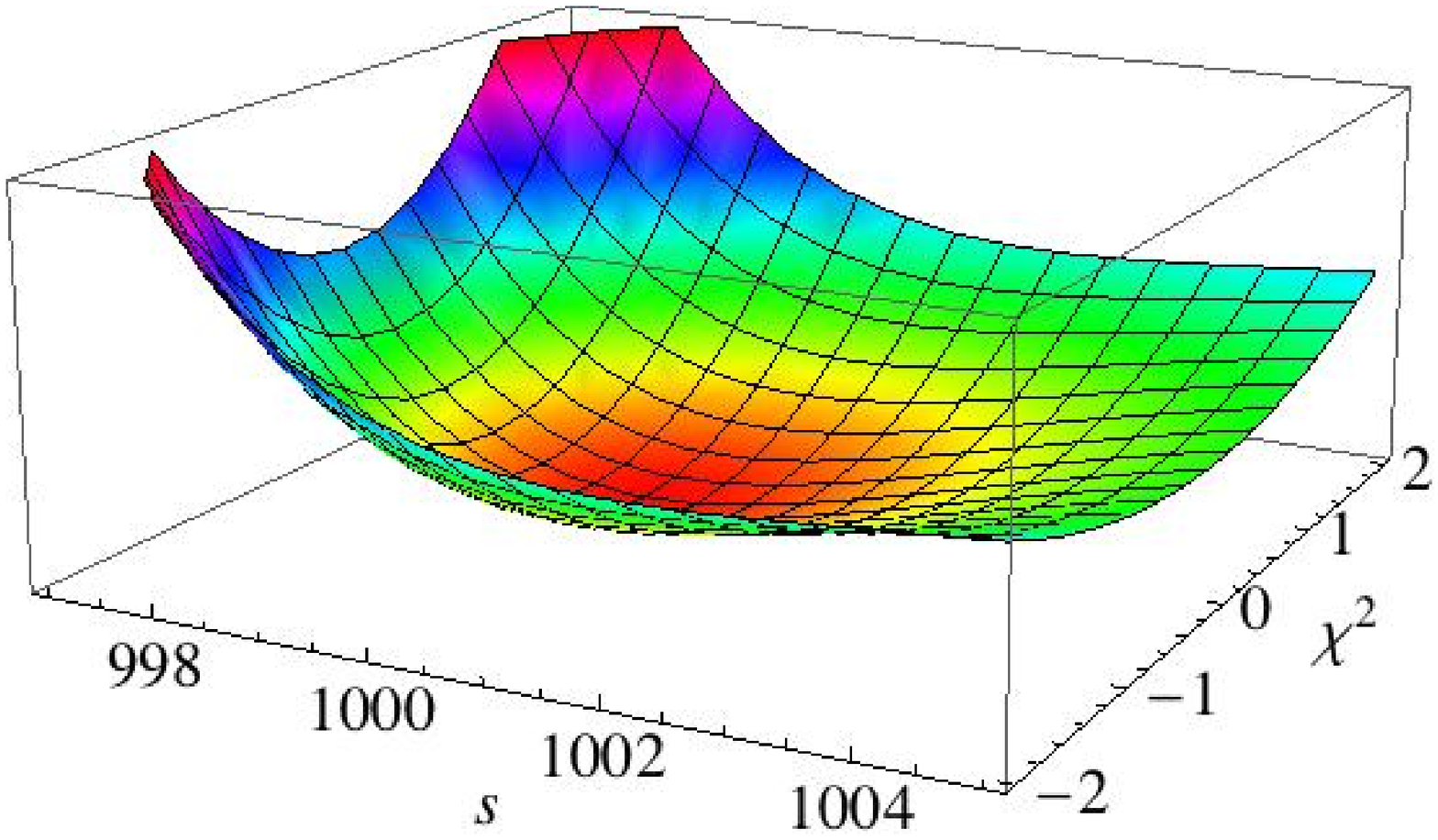} }}
\caption{\emph {Plots of the potential, for the example in Section 5.3 of
    the text, for various slices through field space. The left hand
    image presents the potential as a function of $s$ and $t^1$,
    whereas the right hand image depicts the $s$, $\chi^2$ plane. The
    plots are color shaded as a function of the height of the
    potential. Clearly the vacuum is a
    minimum of the potential in these directions, as confirmed, for all field
    directions, by a calculation of the eigenvalues of the mass
    matrix.}} \label{fig1and2}
\end{figure}

Up to this point, the F-term equations have not depended on the
specific choice of Calabi-Yau three-fold, except through the value of
$h^{1,1}(X)$.  In particular, to discuss the stabilization of the overall
volume and the dilaton, we have not needed the intersection numbers of the
three-fold in any way. To go further, however, and write down the
specific solution for both K\"ahler moduli, one must introduce
this data. We then use the D-term constraint \eqref{snow3}, that is,
\bea \label{dtermconstr} c_1^i ({\cal L})d_{ijk} t^j t^k  =0 \ ,
\eea 
together with the values of $s$ and $\beta_i t^i$ fixed by the
F-terms, to determine the stabilized values of the real parts of the K\"ahler
moduli, $t^{i}$.  To proceed, one must now specify, in addition to the
triple intersection numbers $d_{ijk}$ of the Calabi-Yau three-fold, the
charges $\beta_{i}$ and the explicit anomalous $U(1)$ in the hidden
sector. We take the Calabi-Yau three-fold to be that given in equation
\eqref{egconf}, which has non-vanishing intersection numbers
\begin{equation}
d_{111}=2,~d_{112}=8,~d_{122}=12, ~d_{222}=8
\label{red1}
\end{equation}
as well as those related to the above by symmetry of the indices. We choose the anomalous $U(1)$
in the hidden sector to be associated with the
line bundle 
\begin{equation}
{\cal L}= {\cal O}_{X}(-2,1) \ .
\label{red2}
\end{equation}
 Finally, let
 \begin{equation}
{\beta}= (1,2) \ .
\label{red3}
\end{equation}
 Note that, as required by \eqref{constr}, $\beta_i c_1^i({\cal L}_1)=0$. Having explicitly chosen the Calabi-Yau three-fold, this choice of $\beta_{i}$ corresponds to a specification of the second Chern class of the non-Abelian part of the hidden sector gauge bundle, that is, $c_{2}({\cal{U}})$.  Thus, again, despite the fact that ${\cal{U}}$ does not enter the calculation in stages 2 and 3, the conditions required to solve for the vacuum put further constraints on the choice of ${\cal{U}}$. Given these
choices, \eqref{dtermconstr} tells us that \bea t^1 = (1 + \sqrt{5})
t^2 \;.  \eea Using the fact that $\beta_i t^i=100$ and the value of
$\beta_i$ in \eqref{red3}, we find
\begin{equation}
t^1 = 61.8~, \quad t^2 = 19.1~\ .
\label{red4}
\end{equation}
As stated in the previous subsection, the vacuum we are describing has
vanishing vevs for the axionic components of the K\"ahler modulus and the dilaton stabilized by the F-terms. The remaining
axion, associated with the K\"ahler modulus fixed by the D-term, is a
``flat direction'' of the potential -- as is required by the fact that
it will be ``eaten'' in the process of the associated anomalous gauge
boson becoming massive. Putting everything together, we have shown
that in this example the vevs of the moduli are
\begin{equation}
\langle s \rangle=1000,~\langle \sigma \rangle=0,~\langle t^{1} \rangle=61.8,~\langle t^{2} \rangle=19.1,~\langle \chi \rangle=0 \ .
\label{red5}
\end{equation}
Finally, it is easily demonstrated that the vacuum presented here has
a positive definite mass squared matrix for all fields. That is, it corresponds to a
supersymmetric {\it minimum} of the potential and not merely a saddle
point. Some plots of the potential for various slices through field
space are presented in Figure \ref{fig1and2}.
We emphasize that stage 1 also results in a minimum of the potential
for the $h^{2,1}=50$ complex structure moduli. Thus, this vacuum is a
true minimum of the full theory. The minimum is Minkowski at the
perturbative level. However, the non-perturbative effects induce a small non-vanishing superpotential in the vacuum -- as can be
verified by substituting the vevs \eqref{red5} into the superpotential
\eqref{mrw} -- resulting in a shallow AdS vacuum at the end of stage 3.

There are various important consistency conditions that this example
should, and does, satisfy. For example, all of the expansion parameters
of the four-dimensional theory can be computed and are sufficiently
small that the approximations used in the analysis are valid. In
addition, the second Chern class of the non-Abelian part of the hidden
sector gauge bundle is such that it satisfies the Bogomolov bound for
the stabilized values of the K\"ahler moduli, whatever the rank of
that bundle may be. This is required for this Chern class to be
consistent with the existence of a supersymmetric bundle stabilizing
the complex structure moduli.

\section{Summary, conclusions, and future directions}

The goal of this paper is to provide a new stabilization scenario for
the {\it geometric} moduli -- that is, the dilaton, complex structure
and K\"ahler moduli -- of smooth heterotic compactifications. Our
approach has several novel features. These include using the natural
constraints arising in a heterotic theory -- namely the holomorphy and
slope-stability of the visible and hidden sector gauge bundles -- to
perturbatively stabilize most of the moduli. The three stages of this
scenario are as follows.

First, in {\it stage 1} the complex structure moduli are stabilized by
the presence of a vector bundle which is holomorphic only for an
isolated locus in complex structure moduli space. This geometric
mechanism can, in concrete examples, be described by explicit F-term
contributions to the effective potential. In this approach, the
stabilization of the complex structure is achieved without introducing
flux. As a result, the compactification remains a Calabi-Yau
three-fold, and hence we are able to retain a considerable
mathematical toolkit for analyzing such geometries.

In {\it stage 2}, it is possible to use the remaining perturbative
condition of slope-stability to restrict the dilaton and K\"ahler
moduli. This corresponds to partial D-term stabilization of these
fields. We demonstrate that the presence of these D-terms is highly
constraining to the effective theory. In particular, the D-terms used
in stage 2 are associated with gauging various linear combinations of
axions. Any non-perturbative superpotential must be consistent with
this.

Finally, in {\it stage 3}, we introduce more familiar non-perturbative
effects such as gaugino condensation and membrane instantons. However,
a significant feature of our scenario is that the presence of the
D-terms in stage 2 highly constrains the possible non-perturbative
effects in stage 3. We prove a ``no-go" result -- namely, if only one
linear combination of the K\"ahler moduli and dilaton is left
unstabilized in stage 2, there exists no AdS vacuum of the full theory
including non-perturbative effects. However, it is possible to avoid
this no-go result by allowing two free moduli to remain at the end of
stage 2. We demonstrate explicitly that, in this case, the
non-perturbative mechanisms of stage 3 can complete the stabilization.

A crucial aspect of this scenario is that, at the end of stages 1 and
2, the resulting moduli space of vacua is supersymmetric and
Minkowski. That is, the unstabilized fields have no potential and the
classical cosmological constant is zero. As a result, this scenario
does not suffer from a need to ``fine-tune" the perturbative potential
to be small, as arises in some ``KKLT"-like scenarios.

It should be noted that while the geometric and effective field theory
arguments given in this paper are complete, the results presented here
are still a ``scenario'' since we have not provided a {\it complete}
example of all three stages on a single Calabi-Yau three-fold. To find
such an example, and to couple it to realistic particle physics in the
visible sector, would be an important step forward in heterotic model
building. A search for such geometries and vacua is currently
underway. This will be the subject of future work \cite{usnewfull}.

Finally, it is essential to stabilize the remaining compactification
moduli not considered in this paper -- namely, the vector bundle
moduli, counted by $h^1(V\otimes V^*)$. Potential mechanisms for such
stabilization are already evident in the proceeding sections. While
stages 1 and 2 are largely independent of these moduli, the
non-perturbative effects considered in stage 3 are inherently bundle
moduli dependent. Specifically, the pre-factors of the superpotential
contributions of both gaugino condensation and membrane instantons,
\eref{gaugeino1} and \eref{membrane1} respectively, manifestly depend
on the bundle moduli. These pre-factors are complicated, manifold
dependent polynomials in these moduli. Their specific form,
particularly the bundle moduli dependent pfaffians associated with
membrane instantons, has been studied in
\cite{c1}. We hope to explore this
structure and the stabilization of the vector bundle moduli in future
work.
\section*{Acknowledgments}

L.A. and B.A.O. are supported in part by the DOE under contract number No. DE-AC02-76-ER-03071 and the NSF under RTG DMS-0636606 and NSF-1001296. James Gray would like to thank the University of Pennsylvania for hospitality while part of this research was completed.

\appendix
\section*{Appendix}
\section{Complex structure moduli and NS flux}
In this Appendix, we discuss the complex structure dependent heterotic
superpotential $W$ generated by Neveu-Schwarz flux. This topic lies
somewhat outside our main line of development.  However, as we will
see, the negative results presented here can be seen, in part, as the
motivation for studying the alternative moduli stabilization
mechanisms in heterotic theories discussed in this paper. The analysis
of this Appendix assumes one can continue to work on a Calabi-Yau
three-fold despite the introduction of NS flux \cite{Cyrill}.

The heterotic NS superpotential fixes the complex structure. However, it
also destabilizes the other moduli, specifically the K\"ahler moduli
and the dilaton. Overall stabilization of the model requires adding
non-perturbative effects, such as gaugino condensation or instantons. For this to work, the non-perturbative potential and the flux
potential have to be comparable in size so that the perturbative
runaway can be balanced by the non-perturbative effects. Since
non-perturbative effects are exponentially suppressed, one way to
achieve this is by having a small flux superpotential, similar to what is
required for the KKLT scenario in type IIB theories. We would like to analyze
whether such a small flux superpotential is possible for heterotic NS
flux. Given that the parameters in $W$ are quantized flux, this is by
no means obvious. In type IIB, this can be achieved by an appropriate
``tuning'' of the integer NS and RR flux, but in the heterotic case
only NS flux is available.

We begin by introducing the projective complex structure fields ${\cal Z}^A=({\cal Z}^0,{\cal Z}^a)$. The heterotic NS flux potential then takes the form
\begin{equation}
 W=n_A{\cal Z}^A-m^A{\cal F}_A\; ,
\end{equation}
where ${\cal F}_A=\partial{\cal F}/\partial{\cal Z}_A$ are the derivatives of the pre-potential ${\cal F}$ and $n_A$, $m^A$ are flux integers. We would like to study this superpotential in the large complex structure limit where the pre-potential is given by
\begin{equation}
 {\cal F}=\frac{\tilde{d}_{abc}{\cal Z}^a{\cal Z}^b{\cal Z}^c}{6{\cal Z}^0}\; ,
\end{equation}
with $\tilde{d}_{abc}$ the intersection numbers of the mirror Calabi-Yau manifold. In terms of the physical fields $Z^a={\cal Z}^a/{\cal Z}^0$, the associated flux superpotential in the large complex structure limit reads
\begin{equation}
 W=n_0+n_aZ^a-\frac{1}{2}\tilde{d}_{abc}m^aZ^bZ^c+\frac{1}{6}m^0\tilde{d}_{abc}Z^aZ^bZ^c\; .
\end{equation} 
It is useful to split the fields into their real and imaginary parts as $Z^a=\zeta^a+iz^a$. Further, we introduce the quantity $\kappa=\tilde{d}_{abc}z^az^bz^c$, which is proportional to the volume of the mirror manifold and, hence, should be large in the large complex structure limit, as well as its derivatives $\kappa_a=\tilde{d}_{abc}z^bz^c$ and $\kappa_{ab}=\tilde{d}_{abc}z^c$. 

What we would like to study, for now at large complex structure, is whether $W$ can be made small at a supersymmetric point, that is, at a solution of the F-equations $W_a\equiv\partial W/\partial Z^a=0$~\footnote{While these are the global F-equations, the local ones only differ by a term proportional to $W$ which is negligible if $W$ is small. Hence, absence of solutions with small $W$ at the global level implies their absence at the local level.}. The imaginary parts of the F-equations read
\begin{equation} 
{\rm Im}(W_a)=\kappa_{ab}(m^0\zeta^b-m^b)=0\; . \label{zetasol}
\end{equation}
It turns out that the matrix $\kappa_{ab}$ must be non-singular. This follows because the K\"ahler metric for the complex structure moduli, given by
\begin{equation}
 K_{ab}=-\frac{3}{2}\left(\frac{\kappa_{ab}}{\kappa}-\frac{3}{2}\frac{\kappa_a\kappa_b}{\kappa^2}\right)\; ,
\end{equation}
must be non-singular. Consequently, we can solve Eq.~\eqref{zetasol} for $\zeta^a=m^a/m^0$. (Here we can assume that $m^0$ is non-vanishing. Otherwise, all fluxes except $n_0$ are forced to zero and no moduli are fixed.) Inserting this result into the real parts of the F-equation gives
\begin{equation}
 {\rm Re}(W_a)=n_a-\frac{1}{2m^0}\tilde{d}_{abc}m^bm^c-\frac{m^0}{2}\kappa_a\; , \label{Wa}
\end{equation} 
while the imaginary part of the superpotential can be written as
\begin{equation}
{\rm Im}(W)=n_az^a-\frac{1}{2m^0}\tilde{d}_{abc}z^am^bm^c-\frac{m^0}{6}\kappa\; .
\end{equation}
Multiplying Eq.~\eqref{Wa} with $z^a$ and subtracting this from ${\rm Re}(W)$, one easily finds
\begin{equation}
 {\rm Im}(W)=\frac{m^0}{3}\kappa\; .
\end{equation} 
Since $m^0$ is a flux integer and $\kappa$ needs to be large in the large complex structure limit, this result implies that $|W|$ cannot be made small. Hence, the heterotic flux superpotential is always large in the large complex structure limit.

What happens if we depart from the large complex structure limit? In
this case, the pre-potential ${\cal F}$ becomes a complicated function
which was first computed for specific examples in
Refs.~\cite{Candelas:1990rm,Candelas:1990qd}. While a general analysis
covering the complete moduli space is not straightforward, we have
looked at another limit, namely the region of moduli space near the
conifold point. We have also performed a simple computer scan of the
models of Refs.~\cite{Candelas:1990rm,Candelas:1990qd} and we again find
that $|W|$ cannot be made small at a supersymmetric vacuum. In
conclusion, although we cannot show in general that $|W|$ is large for
vacua away from the large complex structure limit, we have been
unable to find any counterexamples.


\end{document}